\documentclass[12pt]{article}
\usepackage{a4}
\newcommand{\beao}{\begin{eqnarray*}}
\newcommand{\eeao}{\end{eqnarray*}}
\newcommand{\be}{\begin{equation}}
\newcommand{\ee}{\end{equation}}
\newcommand{\bea}{\begin{eqnarray}}
\newcommand{\eea}{\end{eqnarray}}
\newcommand{\beq}{\begin{eqnarray}}
\newcommand{\eeq}{\end{eqnarray}}%vesion with corrections 070221
\newcommand{\nn}{\nonumber}

\newcommand{\Ref}[1]{(\ref{#1})}

\usepackage{epsfig}
\begin{document}
\title{Observables for model-independent detections of    $Z'$  boson at the ILC }

\author{V.~Skalozub $^{a}$\thanks{E-mail: skalozubv@daad-alumni.de}~
 and~I. Kucher $^{b}$\thanks{E-mail: kucherinnaua@gmail.com}
 \\
  {\small\textit{ $^a$ Dnipropetrovsk National University, 49010 Dnipropetrovsk, Ukraine}}\\
% {\small\textit{ Campus UAB, 08193 Bellaterra, Barcelona, Spain}} \\
 {\small\textit{ $^b$ Dnipropetrovsk National University, 49010 Dnipropetrovsk, Ukraine}} \\}
\date{\small}
\maketitle
\begin{abstract}
The  integral observables   for model-independent detections of  Abelian   $Z'$ gauge boson in $e^+ e^- \to \mu^+ \mu^- (\tau^+ \tau^-)$ process with unpolarized beams at the ILC energies are proposed. They are based on the differential cross-section of deviations from the standard model predictions calculated with a low energy effective Lagrangian  and taking into  consideration  the relations between the $Z'$ couplings to the  fermions derived already.   Due to these relations, the cross-section    exhibits angular distribution  giving a possibility for introducing one- or two parameter observables which effectively fit  the mass $m_{Z'}$, the  axial-vector  $a_{Z'}^2$  and the product  of vector couplings  $v_e v_\mu (v_e v_\tau)$. Determination of
the basic $Z'$ model  is discussed. Comparison with other results and approaches  is given.
\end{abstract}

\section{Introduction}
Searching for new heavy particles beyond the energy scale of the
standard model  (SM) is one of the  main goals of modern  high
energy physics. Nowadays it is established on the base of
experimental data accumulated at hadron colliders such as
Tevatron and the LHC. As it was planned beforehand,
distinguishable important  discoveries of these experiments will
be further investigated in details  at the ILC which  will have
energies of   $\sim 500 - 1000$ GeV  in the center-of-mass of
beams but  much better precision of measurements due to point-like
structure of leptons and  experiments with polarized initial and
final fermions.

One of  expected heavy particles beyond the SM is  $Z'$ gauge
boson  which is related with an additional $\tilde U(1)$ group. It
enters as a necessary element numerous GUT  models like $SO(10)$,
$E_6$ as well as superstrings, extra dimensions, etc. Detailed
description of the $Z'$ is given in
\cite{Rizzo89}-\cite{Langacker2008}. Searches for this particle
have been established already within the LEP data   in either
model-dependent \cite{Langacker09}  or model-independent
\cite{Gulov2010} approaches,  and the Tevatron  data
\cite{Kozhushko2011}, \cite{Langacker2011}.  Modern
model-dependent measurements constrain    that the mass  $m_{Z'}$  to be
larger than $2.5 - 2.9$ TeV \cite{ATLAS}, \cite{CMS}. So, at the
ILC experiments the $Z'$  will be investigated as a virtual state.

 At present about hundred $Z'$ models  are discussed in the literature.
 In model-dependent searches established, only  the most popular ones
 such as $LR$, $ALR$, $\chi$, $\psi$, $\eta$, B - L, $ SSM,$ have been investigated
  and the particle  mass  estimated. These models are also used as benchmarks in
   introducing effective observables for future experiments at the ILC \cite{Pankov2010}, \cite{Langacker2013}.
   Analysis of the RS model of strong gravity in $e^+ e^-$ annihilation into leptons see, for example, in \cite{Kisselev2007}.
   Role beam polarizations in detecting of various anomalous particle couplings  is discussed for many years in the literature.
    One of beginning papers is \cite{Pankov1996}. It worth also mentioning that most investigations devoted to model-dependent
    searches at the ILC deal with the  polarized beams and corresponding observables are introduced.

 On the other hand, recent  studies  of perspective variables for identification of the $Z'$ models  \cite{Langacker2013},
 in particular, came to conclusion  that, as complementary way, a model-independent approach is very desirable.
  An important feature of this method is that not only the $Z'$  mass  but also the couplings to the SM fermions are
    unknown parameters which must be fitted in experiments.  Estimations of couplings can be further used in specifying
    of the basic $Z'$ model. Usually, the couplings are considered as independent arbitrary numbers. However, this is not
     the case and they are  correlated parameters, if  some natural requirements,  which this model has to satisfy, are assumed.
 For instance, in most cases we believe that the basic model is renormalizable one. Hence, correlations follow and the amount of  free
  low energy    parameters  reduces.  Moreover, the correlations between couplings influence kinematics of the processes that gives a possibility for
 introducing the specific observables which uniquely pick out the virtual state of interest --  $Z'$ boson in our case.
  The noted additional requirement  assumes searching for new particles within the class of renormalizable models.
   In other aspects the models are not specified. In what follows, we will say  "model-independent approach" in the case
   when either the mass or the couplings must be fitted. Clearly that different correlations fix  coupling properties common
   to the  specific classes of models. Such type analysis is in  between the customary
     model-dependent method, when all the couplings are fixed and only the mass $m_{Z'}$ is free parameter, and  model-independent searches assuming complete independence of couplings describing new physics.

In the present paper we  search for  the Abelian $Z'$ boson
       coming from the extended renormalizable model.  We also assume that there is only one additional heavy particle relevant at considered energies. There are numerous models of such type. In particular,
        most of  $E_6$ motivated models and mentioned above ones belong to this class. In general,
        the requirement of renormalizability admits two sets of correlations between the
 low energy couplings \cite{Gulov2000}. First is  used in what follows (Eq.\Ref{grgav}).
  Second corresponds to a massive neutral vector particle interacting with left-handed
  fermion species, only. It covers other class of extended models
  (see, for example, \cite{Lynch2000}). Searches for this type particle require
   other observables and separate analysis. For more details see Refs.\cite{Gulov2000},
  \cite{ChiralZ'}. In what follows,  we say $Z'$ boson for the  Abelian  one. We also assume,
  as usually, that the SM is the subgroup of the extended group and therefore no interactions
of the  type $Z Z' W^+ W^-$ appear  at a tree-level.

We  apply the model-independent search for the $Z^\prime$ by
analyzing  the deviations  of the differential
  cross-sections for the annihilation process $e^{+}e^{-} \to \mu^{+}\mu^{-} (\tau^+ \tau^-)$ from the SM predictions considered at center-of-mass energies 500 - 1000 GeV. We introduce new observables, $ A(E, m_{Z'})$ \Ref{asyma2},   giving a possibility for estimating both the axial-vector coupling of the $Z'$ to the SM fermions $a_{Z'}$ and the mass $m_{Z'}$, and the  observable $V(E, m_{Z'})$ \Ref{V}, for fitting the products of vector couplings $v_e v_\mu$, $v_e v_\tau$ and the mass $m_{Z'}$.  Our analysis is carried out within  the effective low energy Lagrangian introduced in \cite{Sirlin1989},\cite{Degrassi1987} which describes interactions of the $Z'$ with the SM fermions. It was used already for the $Z'$ searches at the LEP experiments. Detailed description of it and the obtained results are presented in the review \cite{Gulov2010}.
Here we apply this approach with  modifications necessary in searching for the $Z'$ at the ILC. At giving energies and expected particle masses, distinguishable  properties of the factors at couplings entering the cross-section   are observed that gives a possibility for introducing noted observables. Their values can be used  in subsequent determination of the  basic  $Z'$ model. Moreover, the ratio of $ A(E, m_{Z'})$ (or $ V(E, m_{Z'})$) taken at different energies  depends  on the $  m_{Z'}$, only and may be used as new observables for model-independent estimation of it. 

The paper is organized as follows. In next section we adduce necessary information on the approach used and write down the differential cross-section of deviations  from the SM due to the $Z'$ contributions. In section 3 we analyze for different energies and mass $ m_{Z'}$ the factors at  the couplings entering this cross section. On the base of these considerations new integral observable   dependent on the axial-vector coupling $a_{Z'}$ and  mass $ m_{Z'}$ is introduced.
In section 4 the observable for model-independent estimate of $ m_{Z'}$ is proposed.
In section 5 the observable for detecting the product of the vector couplings and the mass is introduced.
The last section is devoted to  the discussion  of the results obtained and  comparison with other approaches.
In Appendix 1 explicit expressions for the cross-sections are adduced. In Appendix 2 as application the model-independent
discovery reach for the mass $m_{Z'}$ is estimated by using the ratio of observable $V(E,  m_{Z'})$ taken at different energies. In Appendix 3 we calculate the discovery reach following from the observable $A(E,  m_{Z'})$ with accounting for the values  $a_{Z'}^2$ estimated from the data set of  LEP experiments.
\section{Cross-section for the $Z'$ detections}
At low energies,  $Z'$ boson can manifest itself as  virtual
intermediate state through
the couplings to the SM fermions and scalars. Moreover, the $Z$ boson couplings are also
modified due to a $Z$--$Z'$ mixing. As it is known (see reviews
\cite{Leike1999,Langacker2008},\cite{Gulov2010},), significant signals beyond the SM can be
inspired by the couplings of renormalizable type. Such couplings
can be described by adding new $\tilde{G}(Z')$-terms to the
electroweak covariant derivatives $D^\mathrm{ew}$ in the
Lagrangian \cite{Sirlin1989}, \cite{Degrassi1987}
\begin{eqnarray} \label{Lf}
L_f &=& i \sum\limits_{f_L} \bar{f}_L \gamma^\mu \left(
\partial_\mu - \frac{i g}{2} \sigma_a W^a_\mu - \frac{i g'}{2}
B_\mu Y_{f_L} -
\frac{i \tilde{g}}{2}\tilde{B}_\mu \tilde{Y}_{f_L}\right) f_L \\
\nonumber &+& i \sum\limits_{f_R} \bar{f}_R \gamma^\mu \left(
\partial_\mu  - i
g' B_\mu Q_{f} - \frac{i \tilde{g}}{2}\tilde{B}_\mu
\tilde{Y}_{f_R}\right) f_R, \\ \label{Lscal} L_\phi &=& \left|
\left(
\partial_\mu - \frac{i g}{2} \sigma_a W^a_\mu - \frac{i g'}{2}
B_\mu Y_{\phi}  - \frac{i \tilde{g}}{2} \tilde{B}_\mu
\tilde{Y}_{\phi} \right) \phi\right|^2,
\end{eqnarray}
where summation over all the SM left-handed fermion doublets,
leptons and quarks, $f_L = {(f_u)_L, (f_d)_L}$, and the
right-handed singlets, $f_R = (f_u)_R, (f_d)_R $, is understood.
In these formulas, $g, g', \tilde{g}$ are the charges associated
with the $SU(2)_L, U(1)_Y,$ and  the $Z'$ gauge groups,
respectively, $\sigma_a$ are the Pauli matrices, $Q_f$ denotes the
charge of $f$ in positron charge units, $Y_{\phi}$ is the $U(1)_Y$
hypercharge, and $Y_{f_L}= -1$ for leptons and 1/3 for quarks. In
  case of Abelian $Z'$, the   $\tilde{Y}_{f_L}= \tilde{Y}_{f_L} \mathrm{diag}(1, 1)$
and $\tilde{Y}_{\phi} = \tilde{Y}_{\phi} \mathrm{diag}(1,
1)$ are diagonal $2\times 2$ matrices with corresponding coupling factors.
These generators
 do not influence the $SU(2)_L$ symmetry.

The
$Z$--$Z'$ mixing angle $\theta_0$ is determined by the coupling
$\tilde{Y}_\phi$ as follows
\begin{equation}\label{2}
\theta_0 =
\frac{\tilde{g}\sin\theta_W\cos\theta_W}{\sqrt{4\pi\alpha_\mathrm{em}}}
\frac{m^2_Z}{m^2_{Z'}} \tilde{Y}_\phi
+O\left(\frac{m^4_Z}{m^4_{Z'}}\right),
\end{equation}
where $\theta_W$ is the SM Weinberg angle, and
$\alpha_\mathrm{em}$ is the electromagnetic fine structure
constant. Although the mixing angle is a small quantity of order
$m^{-2}_{Z'}$, it contributes to the $Z$-boson exchange amplitude
and cannot be neglected in general. There are precision constrains on its
value, coming, in particular, from the LEP1 experiments. It is one of the main
parameters of the $Z'$ physics.  It will be systematically accounted for in what follows.

Below, we will  use the $Z'$ couplings to the vector
and axial-vector fermion currents defined as
 \begin{equation} \label{av} v_f =
\tilde{g}\frac{\tilde{Y}_{L,f} + \tilde{Y}_{R,f}}{2}, \qquad a_f =
\tilde{g}\frac{\tilde{Y}_{R,f} - \tilde{Y}_{L,f}}{2}.
\end{equation}
The Lagrangian (\ref{Lf}) leads to the following interactions
between the fermions and the $Z$ and $Z'$ mass eigenstates:
\begin{eqnarray}\label{ZZplagr}
{\cal L}_{Z\bar{f}f}&=&\frac{1}{2} Z_\mu\bar{f}\gamma^\mu\left[
(v^\mathrm{SM}_{fZ}+\gamma^5 a^\mathrm{SM}_{fZ})\cos\theta_0
%+\right.\nonumber\\&&\quad\left.
+(v_f+\gamma^5 a_f)\sin\theta_0 \right]f, \nonumber\\
{\cal L}_{Z'\bar{f}f}&=&\frac{1}{2} Z'_\mu\bar{f}\gamma^\mu\left[
(v_f+\gamma^5 a_f)\cos\theta_0
%-\right.\nonumber\\&&\quad\left.
-(v^\mathrm{SM}_{fZ}+\gamma^5
a^\mathrm{SM}_{fZ})\sin\theta_0\right]f,
\end{eqnarray}
where $f$ is an arbitrary SM fermion state; $v^\mathrm{SM}_{fZ}$,
$a^\mathrm{SM}_{fZ}$ are the SM couplings of the $Z$-boson.

As it  occurs, if the extended model is renormalizable, the relations between couplings hold
(see \cite{rgr}, \cite{Gulov2000}, \cite{Gulov2010}):
\begin{equation} \label{grgav}
v_f - a_f= v_{f^*} - a_{f^*}, \qquad a_f = T_{3f}
\tilde{g}\tilde{Y}_\phi.
\end{equation}
Here $f$ and $f^*$ are the partners of the $SU(2)_L$ fermion
doublet ($l^* = \nu_l, \nu^* = l, q^*_u = q_d$ and $q^*_d = q_u$),
$T_{3f}$ is the third component of weak isospin. They also can be
derived
 by imposing  the requirement of  invariance of the SM Yukawa term
 with respect to the $\tilde{U}(1)$ gauge transformations \cite{Gulov2001RGR}.
Therefore the relations \Ref{grgav} are independent of the number
of scalar field doublets.

The couplings of the Abelian $Z'$ to the axial-vector fermion
current have a universal absolute value proportional to the $Z'$
coupling to the scalar doublet. Then, the $Z$--$Z'$ mixing angle
(\ref{2}) can be determined by the axial-vector coupling. As a
result, the number of independent couplings is significantly
reduced. Because of the universality we will omit the subscript
$f$  and write $a$ instead of $a_f$.

We assume no new light particles. The relations \Ref{grgav} could change
essentially if the SM has to be modified at energies below the
$Z'$ mass. Thus, we suppose no supersymmetry below the $Z'$
decoupling scale.

Although the relations (\ref{grgav}) were derived for effective low-energy
parameters, they nevertheless also hold at tree-level in a wide
class of known models containing the Abelian $Z'$ (see \cite{Gulov2010}). They also
fulfil  for the case of the Two-Higgs-Doublet SM.
 This is  the reason to call them
 model-independent ones.
The correlations (\ref{grgav}) essentially influence the kinematics of scattering processes and
give a possibility to  uniquely detect the mass and couplings of the virtual $Z'$ boson state.

Let us  consider the process $e^+e^-\to l^+l^-$ ($l=\mu,\tau$)
with the non-polarized initial and final fermions.  Two classes of
diagrams have to be taken into consideration. The first one
includes the pure SM graphs. This part should be estimated as
accurate as possible. The second class  includes heavy $Z'$ boson
as the virtual state described by the effective Lagrangian
\Ref{ZZplagr} and the scalar particle  contributions. We assume
that $Z'$ is decoupled and not excited inside loops  at the ILC
energies. The tree-level diagram $e^+e^-\to Z'\to l^+l^-$ defines
a leading contribution to the cross-section. It is enough to take
into account this diagram to estimate the $Z'$ signals. The
cross-section includes the contribution of the interference of the
SM amplitudes with the $Z'$ exchange amplitude (having the order
$\sim a^2, v_f a $) and the squared of the latter one (of the
order $\sim a^4, v_f^4$). Since the couplings of the $Z'$ are
small, the last contribution  can be neglected at far from
resonance  energies.
 In our calculations,  radiative corrections to the $Z'$-exchange diagram are
incorporated   in the improved Born
approximation. This mainly influences the values of couplings  at high energies and is sufficient
for applied analysis.

Within theses assumptions, the deviation of the  differential cross-section for the process
$e^+e^-\to\mu^+\mu^-$ ($\tau^+\tau^-$) can be written   in the form (see Eq. (27) in \cite{Gulov2010})
\begin{eqnarray}\label{5}
\Delta \sigma(z)=\frac{d\sigma}{dz}-\frac{d\sigma^\mathrm{SM}}{dz} =&&
f^{\mu\mu}_1(z)\frac{a^2}{m_{Z'}^2} + f^{\mu\mu}_2(z)\frac{v_e
v_\mu}{m_{Z'}^2}
+ \nonumber\\
 +&& f^{\mu\mu}_3(z)\frac{a
v_e}{m_{Z'}^2} + f^{\mu\mu}_4(z)\frac{a v_\mu}{m_{Z'}^2}.
\end{eqnarray}
Here, $z = cos \theta$ is the cosine of scattering angle $ \theta$. Eq. (\ref{5}) is our definition of the $Z'$ signal.

This cross-section
accounts for the relations  \Ref{grgav} through the known dimensionless functions
$f_1(z)$, $f_3(z)$, $f_4(z)$, since the coupling $\tilde{Y}_\phi$
(the mixing angle $\theta_0$) is substituted by the axial-vector coupling
 $a$ which is universal parameter.  Usually, when a four-fermion effective Lagrangian is
applied to describe physics beyond the SM,
this dependence on the scalar field coupling is neglected at all \cite{Babich2003}, \cite{Langacker2013}.
However, in our case, when we are interested in searching for
signals of the $Z'$ boson on the base of the effective low-energy
Lagrangian (\ref{Lf})--(\ref{Lscal}), these contributions to the
cross-section are essential.

To introduce the  observables of interest we have to investigate the $z$-dependence
of the factors $f_i(z)$, $i$=1,...,4  for a number of proper  energies and possible $Z'$ masses. In Appendix 1
we adduce this functions as well as the expression for the SM case in the chosen  approximation.
\section{Observable for  estimation of  $a^2$ and  $m_{Z'}$}
According to Eq.\Ref{5}, the deviation of the differential
cross-sections is  described by  four factors $f_i(z)$. Let us
investigate their behavior assuming that couplings $a, v_f$ have
the same order of magnitude. In this case the kinematics
properties of $f_i(z)$ can be elucidated.

For definiteness, in Figs. 1, 2 we show the behavior for  energy
$E = 500$ GeV in  the $e^+ e^-$ center-of-mass and the mass
$m_{Z'} = 2500,  3000$ GeV.  Here, some remark is needed. As it
was reported in \cite{ATLAS}, \cite{CMS}, the lower bound on the
mass $m_{Z'}$ obtained from the data on the Drell-Yan process at
the LHC is $m_{Z'} \geq 2.5 - 2.9 $ TeV. It has been estimated
assuming the narrow resonances with the width $\Gamma_{Z'}$ of the
order: $\Gamma_{Z'}/m_{Z'} \sim 0.01$. Similar assumptions were
also used in \cite{Langacker2013} for the analysis of the process
$e^+ e^- \to f \bar{f}$ at the ILC with the goal of introducing
effective observables for the determination of the $Z'$ model. On
the other hand, as it was argued in \cite{Kozhushko2011}, the
resonances with not  small $\Gamma_{Z'}$ are not excluded. They
could considerably decrease the value of the lower bound on the
$m_{Z'}$. Below,  to present the results  we take the ratio
$\Gamma_{Z'}/m_{Z'} \sim 0.1$ (the results  for narrow resonances
are similar).

%
%\begin{center}
%\includegraphics[bb=0 0 260 160
%,width=0.5\textwidth]{eemumu2500b..eps}
%\end{center}
%\noindent\textsf{\textcolor{cl02}{\bf{Fig. 1:}}}
%\mbox{\parbox[t]{0.84\hsize}{\textsf{Fig.1 Behavior of factors $f_1(z)$, $f_2(z)$, $f_4(z)$ for $m_{Z'} = 2500$ GeV and $E = 1000 $GeV}}}
%
%\begin{center}
%\includegraphics[bb=0 0 260 160
%%
%,width=0.5\textwidth]{eemumu1500a.eps}
%\end{center}
%\noindent\textsf{\textcolor{cl02}{\bf{Fig. 1:}}}
%\mbox{\parbox[t]{0.84\hsize}{\textsf{Fig.3 Behavior of factors $f_1(z)$, $f_2(z)$, $f_4(z)$ for $m_{Z'} = 1500$ GeV and $E = 1000 $GeV}}}
%

\begin{center}
\includegraphics[width=0.9\linewidth,keepaspectratio]{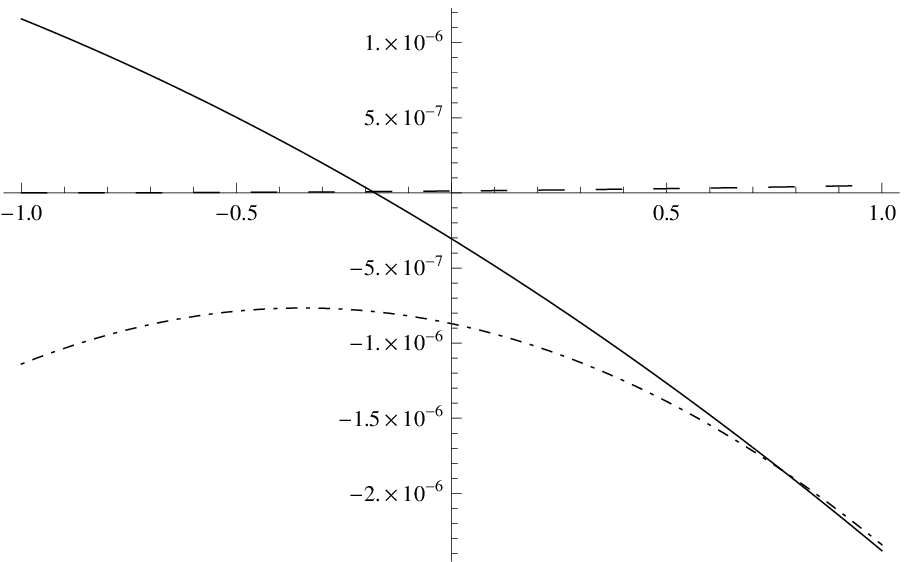}
\end{center}
%\noindent\textsf{\textcolor{cl02}{\bf{Fig. 1:}}}
\mbox{\parbox[t]{0.84\hsize}{\textsf{Fig.1 Behavior of factors $f_1(z)$, $f_2(z)$, $f_4(z)$ for $m_{Z'} = 2500$ GeV,  width $\Gamma_{Z'} = 250 $ GeV
 for  $E = 500 $GeV}}}
\begin{center}
\includegraphics[width=0.9\linewidth,keepaspectratio]{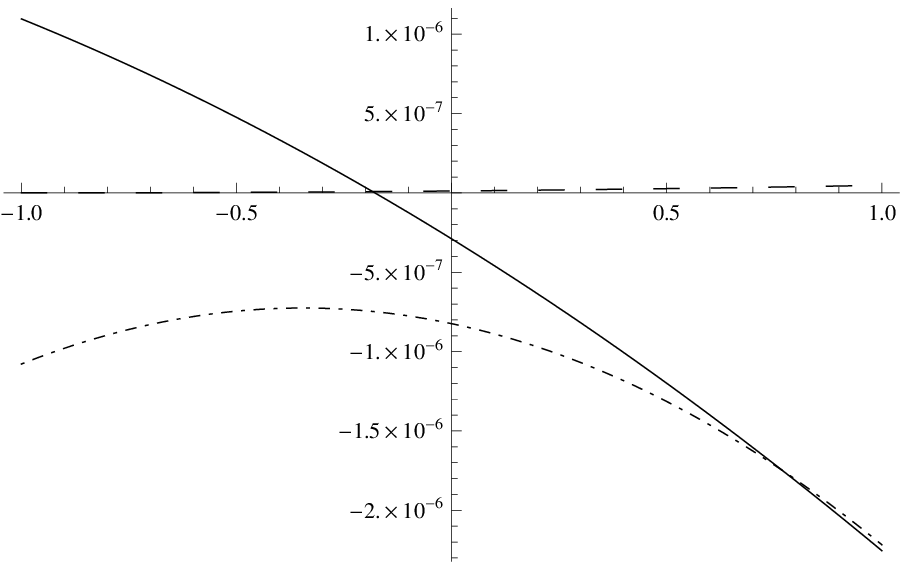}
\end{center}
%\noindent\textsf{\textcolor{cl02}{\bf{Fig. 1:}}}
\mbox{\parbox[t]{0.84\hsize}{\textsf{Fig.2 Behavior of factors
$f_1(z)$, $f_2(z)$, $f_4(z)$ for $m_{Z'} = 3000$ GeV,
 width $\Gamma_{Z'} = 150 $ GeV  for $E = 500 $GeV}}}

 In the plots, the function $f_1(z)$ is presented  as solid line, the $f_2(z)$ is shown as dot-dashed one  and the functions $f_{3,4}(z)$ are shown as dashing line. The $f_{3,4}(z)$ coincide
  at high energies when one can neglect fermion masses. The  function $f_1(z)$  has opposite
   signs for the forward and backward beans. This is in contrast to the factors $f_2(z)$ and $f_{3,4}(z)$.
   The former is negative and the latter - positive one.
Moreover, the factors $f_{3,4}(z)$ are suppressed by two orders of magnitude as compared to the $f_1(z)$ and $f_2(z)$.
Shown angular dependence is typical and takes place in the wide mass interval and for other energies,
for example, $1000$ GeV. The mass interval $1.5 \leq m_{Z'} \leq 4$ TeV  was investigated. This
 behavior makes reasonable   introducing the integral observable which  picks out the contribution
 coming from the first term in Eq. \Ref{5} and consequently the axial-vector coupling $a^2$.

 Really, we can  integrate  $f_2(z)$ in the intervals $(-1 < z < -0.2)$ (where the function $f_{1}(z)$ is positive)
  and $(-0.2 < z < z^*)$ (where $f_{1}(z)$ is negative) and specify the limit $z^*$ in such a way that the
    difference of the integrals turns to zero:

\be \label{za} (\int\limits_{-1}^{-0.2} - \int\limits^{z^*}_{-0.2}) f^{\mu\mu}_2(z) d z = 0. \ee
  Since $f_2(z)$ is sign definite, this point always  exists. At the same time, due to opposite signs
  of $f_1(z)$ in these intervals and sign definiteness  of $f_{3,4}(z)$, the difference is mainly
  determined by the first  term in \Ref{asyma2}. The partial cancelation of the contributions coming
  from $f_{3,4}(z)$  takes place. Although this  is not very essential because of the significant
   suppression of these factors. As a result, the value of the universal coupling constant $a^2$ can
    be estimated with high accuracy. As explicit calculations showed, the upper limit of integration
      equals to $z^* = 0.489$ for a wide interval of both the mass $m_{Z'}$ and beam energies $E$. It
      is also important that the function $f_1(z)$ changes its sign at the point $z = - 0.2$ for all energies and masses investigated.

  On these grounds we introduce the
 observable for model-independent estimating of the $a^2$ and $m_{Z'}$:
 \be\label{asyma2} A(E, m_{Z'}) = \bigl(\int\limits_{-1}^{-0.2} -  \int\limits_{-0.2}^{z*}\bigr)
 \bigl(\frac{d\sigma}{dz}-\frac{d\sigma^\mathrm{SM}}{dz}\bigr)d z. \ee
Here, the lower and upper limits of integration are theoretical bounds. They can be
 substituted by other ones corresponding to actual  set  up of experiments. For example,
 for the lower limit $z_{lower}= -0.9 $,  that is close to the values of measured scattering
 angles, $10^o < \theta < 170^o$, planned for the ILC detectors \cite{ILCdet},  the upper limit is $z^* = 0.406$.
 To complete this section, we adduce the values of the observable \Ref{asyma2} for the number of the  mass  and energy values.

\begin{table}[ht]
\caption{Observable $A(E, m_{Z'})$ for the interval [-1,0.489]}
\begin{tabular}{c c c c c c}
\hline
Energy & $m_{Z'}$ & $\Gamma_{Z'}$ & $f_1(z)$ & $f_{3,4}(z)$ & $f_2(z)$\\[0.5ex]
\hline
500 & 2500 & 250 & $9.03452\cdot 10^{-7}$ & $-9.48604 \cdot 10^{-9}$ & $4.93451\cdot 10^{-10}$ \\
500 & 3000 & 300 & $8.55036\cdot 10^{-7}$ &$ -8.97772\cdot 10^{-9}$&$ 4.67504\cdot 10^{-10}$ \\
\hline
1000 & 2500 & 250 & $1.96179\cdot 10^{-6}$ &$ -2.058\cdot 10^{-8}$ & $-3.67104\cdot 10^{-9}$ \\
1000 & 3000 & 300 &$ 1.3263\cdot 10^{-6}$ &$ -1.39139\cdot 10^{-8}$ &$ -2.4791\cdot 10^{-9} $\\
\hline\hline
\end{tabular}
\end{table}

\begin{table}[ht]
\caption{Observable $A(E, m_{Z'})$ for the interval [-0.9,0.406]}
\begin{tabular}{c c c c c c}
\hline
Energy & $m_{Z'}$ & $\Gamma_{Z'}$ & $f_1(z)$ & $f_{3,4}(z)$ & $f_2(z)$\\[0.5ex]
\hline
500 & 2500 & 250 & $6.98371\cdot 10^{-7}$ & $-7.34139 \cdot 10^{-9}$ & $1.1672 \cdot 10^{-9}$ \\
500 & 3000 & 300 & $6.59927 \cdot 10^{-7}$ &$-6.92419 \cdot 10^{-9}$ & $-8.61322 \cdot 10^{-11}$\\
\hline
1000 & 2500 & 250 & $1.51413 \cdot 10^{-6}$ & $-1.58729 \cdot 10^{-8}$ & $-3.87173 \cdot 10^{-9}$\\
1000 & 3000 & 300 & $1.02365 \cdot 10^{-6}$ & $-1.67314 \cdot 10^{-8}$ & $-2.11544 \cdot 10^{-9}$ \\
\hline\hline
\end{tabular}
\end{table}

In the tables, in first, second and third columns the  energy, mass and  width values (expressed in GeV)
 are given, correspondingly.
In the fourth column the contribution coming from $f_1^{\mu\mu}(z)$ is adduced. In the fifth
 and sixth columns the values of the contributions coming from the factors $f_{3,4}^{\mu\mu}(z), f_{2}^{\mu\mu}(z)$
Eq.\Ref{5} are shown.

As we see,  the contributions  coming from the factors $f^{\mu\mu}_{2,3,4},$ are two order less
compared to the contribution from $f_1^{\mu\mu}(z)$ and can be neglected in the total. In this
way we obtain the two parameter observable for estimation of  $a^2$ and $m_{Z'}$.

Since the contribution of the factor $f_2(z)$ is chosen to be zero, the $A(E, m_{Z'})$ is
determined by two couplings $a^2$ and $a v_\mu$. The  efficiency of the observable
is determined from the relation:
\be \label{effA} \kappa_A = \frac{|f_1^{\mu\mu}|}{|f_1^{\mu\mu}| + |f_{3,4}^{\mu\mu}|}. \ee
Here the quantities $|f_i^{\mu\mu}|, i = 1, 3, 4,$ mark the integrals
\be (\int\limits_{-1}^{-0.2} - \int\limits^{z*}_{-0.2})f_i^{\mu\mu}(z) d z > 0. \ee \nn
From Tables 1, 2 it can be estimated as $\kappa_A = 0.9896 $ for all the given energy and mass values.

The signature of the observable is also important. The coupling $a^2$ is positive and the  integral $|f_1^{\mu\mu}|$
is also positive by construction. So, the positivity of the $A(E, m_{Z'})$ is the distinguishable signal of the $Z'$ boson.

This observable can be used in different ways dependently on the results obtained at the LHC.
Usually, two scenarios are discussed in the literature (see, for example, \cite{Pankov2010}, \cite{Langacker2013}).
First, the $Z'$ is detected and its mass is estimated but the model not. Second, neither the mass $m_{Z'}$ nor the $Z'$ model are
 determined. In the former case, for the given mass $m_{Z'}$, the axial-vector coupling  $a^2$ can be simply
 estimated in a model-independent way within a one parameter fit. In the second case, both of the parameters
 can be found in  two parameter fitting. Since $a^2$ is universal it covers numerous $Z'$ models.
\section{Observable for estimation of  $m_{Z'}$}
Interesting application of the observable $A(E, m_{Z'})$
\Ref{asyma2} is related with  the model-independent determination
of the mass $m_{Z'}$. Really,  the observable \Ref{asyma2}
 includes   the   factor $a^2$ which  is canceled in the ratio $ R^{experim}_A = \frac{A(E_1, m_{Z'})}{A(E_2, m_{Z'})} $ .
 So that the behavior of it can be used in estimating  $m_{Z'}$.  We can  consider
 two cross-sections with close energies $E_1$ and $E_2 = E_1 + \Delta E$  and write
\be \label{Drat} R^{experim}_A = \frac{A(E_1, m_{Z'})}{A(E_2, m_{Z'})} =
1 - \frac{\partial \ln A(E_1, m_{Z'})}{\partial E_1} \Delta E. \ee
As a theoretical curve $R^{theory}_A$ the function $f^{\mu\mu}_1$
from Eq.\Ref{5}  has to  be substituted in Eq.\Ref{Drat} instead
of $A(E_1, m_{Z'})$. This is because  contributions of  all other
form-factors are suppressed in the difference.  As a result, we
obtain the observable  dependent on $m_{Z'}$, only.
Hence, the value of the mass can be estimated by using a standard
$\chi^2$ method. The value  $\Delta E$ can be taken as the
difference  between the closer beam energies of experiments.

One may wonder how the  mass can be estimated
without any information about the Z' boson, which
will depend on the couplings. In particular,  the measurement accuracy of
the Z' mass has to depend on the size of the new-physics signal determined by the
coupling values. However,  information on the coupling is completely removed from $R^{experim}_A$.

Nevertheless, in the developed approach this possibility is realized due to   the following. First, the model-independent analysis
is based on the cross-section of deviations $\Delta \sigma(z)$ \Ref{5}, which tends to zero
if couplings are very small. So that the deviation must be visible. Since the factors $f_i$
 take into account the relations \Ref{grgav}, they uniquely pick out the virtual Z' boson state. Therefore,
we expect that, if some deviation is generated by any other virtual particles, the estimated couplings $a$ and $v$ are to be zero and the mass
very large.
 Second, the dependence on the mass  is a propagator effect which uniquely exhibits
itself though the function $f^{\mu\mu}_1$ after integration over the   interval of $z $ where the contributions of the factors $f_2, f_3, f_4$ are canceled in the total.  This integral is the same for various coupling values. We shall return to this problem in what follows.

The linear approximation used in the l.h.s. of Eq. \Ref{Drat} is sufficient for small $\Delta E$. It can be modified for relatively large $\Delta E$.
In actual investigations, one can start from the mass estimates  and then use the obtained results in  determining  of  $a^2$ by means of  a one parameter fit. This can be used as complementary analysis for the two parameter fitting mentioned above.

\section{Observable for estimation of $v_e v_\mu$ ($v_e v_\tau$)}
The behavior of factors shown in  Fig. 1, Fig. 2 gives also a possibility of introducing the observable for
model-independent determination of the product $v_e v_\mu (v_e v_\tau$).  As we see from the plots and Tables 1, 2, the contributions of the
factors standing at $a  v_{\mu} $and $a v_e$ are  suppressed and can be neglected. To exclude the contribution of the
$a^2$-dependent term we have to integrate the differential cross-section $\Delta \sigma(z)$
\Ref{5} over  $z$ in the interval $(- 1 \leq z \leq z^v )$
and select the upper limit from the requirement
\be \label{zv} \int\limits_{-1}^{z^v} f^{\mu\mu}_1(z) d z = 0. \ee
Hence, we obtain the  observable $V_{e \mu}(E, m_{Z'}) $ for  estimation of  $v_e v_\mu $ (or $v_e v_\tau $):
\be \label{V} V_{e \mu}(E, m_{Z'}) =  \int\limits_{-1}^{z^v}   (\frac{d\sigma}{dz}-\frac{d\sigma^\mathrm{SM}}{dz})d z, \ee
where the limit $z_v$ depends on the energy $E$ and mass $ m_{Z'}$.

Let us adduce the values of $z^v$ and $V_{e \mu}(E, m_{Z'})$  for the number of  energy and mass values.
\begin{table}[ht]
\caption{Upper limit  $z^v$ and the value   $V_{e \mu}(E, m_{Z'})$}
\begin{tabular}{c c c c c c}
\hline
Energy &$ m_{Z'}$ & $\Gamma_{Z'}$ & $z^v$ & $V_{e \mu}\cdot m^2_{Z'}$&$A\cdot V_{e \mu}\cdot m^2_{Z'}$\\[0.5ex]
\hline
500 & 2500 & 250 &  0.567466 & $-1.50333 \cdot 10^{-6}$&$1.65644\cdot 10^{-8}$\\
500 & 3000 & 300 &  0.5675 & $-1.42282\cdot 10^{-6}$&$1.56777\cdot 10^{-8}$\\
\hline
1000 & 2500 & 250 & 0.570118 & $-3.31411 \cdot 10^{-6}$&$3.52717\cdot 10^{-8}$\\
1000 & 3000 & 300 & 0.570115 & $-2.24064 \cdot 10^{-6}$&$2.38447\cdot 10^{-8}$\\
\hline\hline
\end{tabular}
\end{table}
In Table 3, the first three columns show the center-of-mass energy, mass and width, as in Tables 1, 2. In the fourth column the cosine of boundary angles is adduced. In the last two columns the  corresponding values of $V_{e \mu}\cdot m^2_{Z'}$ and the contributions of the factor at the product $a v_\mu$ are presented.
Of course,  these limits can be substituted by other ones according to an experiment set up.
As above,  the contributions  of the  factors  $\sim a v_e,   a v_\mu  $ are negligibly small and can be omitted in the total.

 The  efficiency of the observable $V(E, m_{Z'})$
is determined analogously to the $\kappa_A$ \Ref{effA} according to  the condition
\be \label{effV} \kappa_V = \frac{|f_2^{\mu\mu}|}{|f_2^{\mu\mu}| + |f_{3,4}^{\mu\mu}|}, \ee
where now  the quantities $|f_i^{\mu\mu}|, i = 2, 3, 4,$ mark the integrals over the interval $-1 < z < z^v$.
The
efficiency is estimated as  $\kappa_V = 0.9891$. Again we obtain very efficient observable.

Since the factor $f_2(z)$ is negative, the sign of the observable $V(E, m_{Z'})$ depends on sign of the product $v_e v_\mu$. If this value is positive, we have negatively defined observable.
For this case, the negative sign  is also the distinguishable signal of the virtual $Z'$ boson.

 As a result, we have obtained  the two parameter observable for fitting of the product of vector couplings and the mass $m_{Z'}$.
  In the case of family independence for vector  couplings  $v_e = v_\mu = v_\tau = v$, as it is often assumed,
   $v^2$ can be determined in the model-independent way. The calculation procedures are quite similar to that described for the case of the $a^2$ coupling. Moreover, the positivity  of $A(E, m_{Z'})$ and negativity  of $V(E, m_{Z'})$ is  the distinguishable  signal of the $Z'$ boson.
Analogously to  $A(E, m_{Z'})$ in sect. 4, the observable $V(E, m_{Z'})$ can be used for model-independent estimate of the $Z'$ mass. To demonstrate this ability, in Appendix 2 we derive a discovery reach for   $m_{Z'}$, estimated for some values of $\Delta E$.

The accuracy of possible estimates depends  on both theoretical and experimental uncertainties. The former account for the accuracy of the cross-section calculation, which includes the SM terms and the additional terms coming from the low energy effective Lagrangian Eq.\Ref{Lf}.  The latter  depend on the  precision of measurements. The detailed analysis of that  within  the LEP1 and LEP2 experiments is provided in \cite{Gulov2010}. It also has relevance to considered case.   Here we note that the deviation $\Delta \sigma(z)$ \Ref{5} (and therefore the introduced model independent observables)  with high accuracy can  be related with the standard  variables - the deviation of the total cross-section $\Delta \sigma ^T $ and forward-backward asymmetry $A^{FB}$. It looks as follows:
\be \label{5a} \Delta \sigma (z) = (1 + z^2) \beta + z \eta + \delta (z), \ee
where $\delta (z)$ is the difference between the exact and approximate cross-sections. The parameters $\beta, \eta$ can be calculated as (see \cite{Gulov2010} for details)
\be \label{beta} \Delta \sigma ^T  =  \sigma ^T -  \sigma ^{T,SM} = \frac{8 \beta}{9} + \delta(-1), \ee
\be \label{eta}  \Delta \sigma ^{FB}  =  \sigma ^{FB} -  \sigma ^{FB,SM} = \eta + \delta(0),\ee
where $\delta(-1), \delta(0)$ are the deviations at the   specified $z$.
The forward-backward cross-section can be written in the form:
\be \label{deltaFB} \Delta \sigma ^{FB} = \Delta \sigma ^T A^{FB} + \sigma ^{T,SM} \Delta A^{FB}. \ee
Through these relations the accuracy of measurements of the introduced observables $A(E, m_{Z'}), V_{e \mu}(E, m_{Z'})$ can be related with the accuracy of measurements
of the total cross-section and the forward-backward asymmetry.

As it was estimated  for LEP experiments (see \cite{Gulov2010}),  the  deviation $\delta$ is much less than the systematic error, which includes also theoretical errors, and for the SM  is of the order  2  \%.  So that we assume that  not larger values will be for the ILC.
At considered energies, the contributions of the omitted terms $\sim a^4, (v_e v_\mu)^2$ are estimated
as 0.1 \%. According to data in Tables 1-3, the neglected contributions coming from the factors $f_3, f_4$
are  estimated as 1 - 1.5 \%. Hence, we estimate the theoretical errors  as 3-4 \%.  The accuracy of measurements of the leptonic cross-sections is expected to be high.
Thus, the couplings and the mass $m_{Z'}$  can be precisely  measured either  from the differential cross-sections or from  data on  the total cross-sections.

  The derived values can be used further in determination of the basis  renormalizable $Z'$ model. This procedure depends on the results obtained  at the LHC. If the mass $m_{Z'} $ will be estimated, the couplings $ a^2, v_e v_\mu$ can be determined in the one parameter fits.

\section{Discussion}
We have investigated the process $e^+ e^- \to \mu^+ \mu^- (\tau^+ \tau^-)$ for unpolarized initial and final fermions  at the center-of-mass energies $500 - 1000 $ GeV with the goal of introducing the integral observables for model-independent detections of the  $Z'$ boson. In doing that the relations \Ref{grgav}  have been used.  The account of them considerably reduces the number of parameters which must be fitted in experiments.
 Moreover,  the factors entering the differential cross-section \Ref{5} exhibit features giving a possibility for introducing   the integral  observables  \Ref{asyma2} and   \Ref{V}  dependent mainly on only one coupling $a^2$, or $v_e v_\mu$ ($v_e v_\mu $), correspondingly,  and the mass  $m_{Z'}$. So that all these parameters can be estimated within one- or two  parameter fits. Remind that the coupling $a^2$ is universal according to the relations \Ref{grgav}.

 On the basis of these observables the model-independent estimate of the $Z'$ mass can be done. It  may be of interest if the $Z'$ boson is heavy and could not be discovered at the LHC. At low energies,  the data on the  cross-sections at two different energies  are needed. Then,  the mass $m_{Z'}$ can be found from
 the observable $R^{experim}_A$ \Ref{Drat} related to the observable \Ref{asyma2}, or from the similar observable
  $R^{experim}_{e \mu}$ related to the $V_{e \mu}$ \Ref{V}. To obtain the latter one we have to substitute the function $f^{\mu\mu}_1$ in
   the theoretical expression $R^{theory}_{A}$ by the  $f^{\mu\mu}_2$ from \Ref{5} and make obvious modifications in sect. 4.
   The same can be done for the $\tau$-lepton final states. As a result, the mass  $m_{Z'}$  can be fitted by using
    two different factor functions and therefore there are two  ways of  measuring this parameter.

    It worth to mention that the  observables \Ref{asyma2}, \Ref{V} are specialized mainly for detecting the couplings not the mass $m_{Z'}$. By construction, each  of them is proportional to the  constant  determining the   interaction strength. On the contrary, the dependence  on the mass $m_{Z'}$ is the propagator effect which is described  by smooth functions (see Appendix 1). That is why  the model-independent determination of the $Z'$ mass can not be done with  very good accuracy. This  partially can be  compensated by the number of different fits. In connection with application at the ILC, the main theoretical error of the observables $R_A$ and $R_V$ is related with sufficiently large  interval, $\Delta E = 200 - 300$ GeV,  between beam energies.  For LEP experiment data, where $\Delta E \sim 10$ GeV, they are more reliable.

 To have some ideas about the ability of the observables, in Appendix 2 we estimate  the discovery reach for $m_{Z'}$ followed from  the observable $R_V$ for two values of $\Delta E$ = 300 and 10 GeV. The  value $m_{Z'}^{DRV} $ = 1.5 TeV obtained in the former case is approximately twice less than the lower bound on the mass reported in \cite{ATLAS}, \cite{CMS}. In fact, this could be the consequence  of the estimate roughness  related with  the linear approximation used. So, it  may occur   that two parameter fits including the couplings and the mass are more precise. On the contrary, for $\Delta E$ = 10 GeV $m_{Z'}^{DRV} $ = 12 TeV.  All these  require detailed  analysis and comparisons of the results coming from the $R_A$ and $R_V$  observables. It will be done elsewhere separately.

 In  Appendix 3 we obtain the discovery reach  for $Z'$ with taking into consideration the observable $A(E, m_{Z'})$ \Ref{asyma2} and  the axial-vector coupling $a^2$  (see Eq. \Ref{aLEP} ) estimated from the data set  of LEP experiments and reported in \cite{Gulov2010}:  $m_{Z'}^{DRA} = 4.4 $ TeV. This value follows from the model-independent estimates obtained with accounting for the relations \Ref{grgav}.  It is not much larger than  the low limits on the mass obtained in the model-depended searches for popular models: $m_{Z'} > 2.9 - 3 $ TeV.

 Next what can be verified on the base of the $V(E, m_{Z'})$ observable  is family independence of $v_f$ couplings. Really, the ratio of the observables taken at a fixed energy
 \be \label{ratio} D_v^{\mu \tau} = \frac{V(E, m_{Z'})_\mu}{V(E, m_{Z'})_\tau} = \frac{v_\mu}{v_{\tau}} \ee
 depends on the coupling values and has to be  unit in the case of the family independence. It can be simply checked.

 The observable $ D_v^{\mu \tau}$ can also be used for measuring the couplings $v_{e}, v_{\mu}, v_{\tau}$  in the leptonic processes.   Usually it is believed  \cite{PWG}, \cite{Godfrey}  that an additional information coming from hadronic processes is necessary. This speculation follows from  the fact that in  leptonic cross-sections the couplings enter as the products  $v_{e} v_{\mu} = d_{e \mu} , v_{e} v_{\tau} = d_{e \tau}$. In the considered case, let us assume  that the products $d_{e \mu}$ and $d_{e \tau}$ are measured. Then, the observable \Ref{ratio} equals to: $D_v^{\mu \tau} = d_{e \mu}/d_{e \tau}$. Hence
\be \label{vevmu} v_\mu = v_\tau D_v^{\mu \tau}, ~~v_e = \frac{d_{e \tau}}{v_\tau},\ee
and we can express these couplings in terms of $v_\tau$. Combining this with the results on the Bhabha process $e^+ e^- \to e^+ e^-$,  all the leptonic vector couplings can be measured.

It is essential  that signature of the observables -  positive sign of   $A(E, m_{Z'})$ and  negative sign of $V(E, m_{Z'})$   - is the signal of the Abelian $Z'$ boson.

 The present  approach  can be used  as an additional way   for detecting at the ILC the $Z'$ boson    as well as determining the model which it has to belong.
 Let us consider the case when the $Z'$ resonance state is observed at the LHC and we are interested  in  distinguishing between the  models. This problem is reduced to  distinguishing of model couplings.  For example, in Ref.\cite{Godfrey} the possibility of separating the $\chi$ model coming from the $E_6$ symmetry breaking, LR-symmetric model (LR), Little Higgs model (LH), Simplest LH model (SLH) and KK excitations originating in theories of extra dimensions is discussed. Detailed analysis for the number of expected mass values is given within the  sets of observables and beam polarizations $P_{e^+}, P_{e^-}$. This way is typical for model-dependent analysis.

From the point of view of the present  approach accounting for the  relations  \Ref{grgav} or other ones corresponding to the  chiral Z' boson (see for details \cite{Gulov2010}, \cite{Lynch2000}),  select  some classes of  models. According this classification, the $\chi$ and LR models satisfy  the relations \Ref{grgav} and can be analyzed with the observables considered. The LH and SLH models correspond to the effective theory which is not renormalizable. So, the couplings of these model do not fit these relations. The same concerns the KK model. The $Z'$ models investigated in \cite{PWG} satisfy  the relations \Ref{grgav} and can also be analyzed by means of the introduced observables.

 Then, the found values of the  couplings  can be compered with the values  for the  specific renormalizable $Z'$ models. As a result,  the number  of the perspective   candidates can be  considerably reduced. This  is very  important because the identification reach for the $Z'$ models at the LHC is estimated as  $m_{Z'} \leq 2.2 - 2.3$ TeV whereas  the nowadays model-dependent lower bound is $\sim 2.5 - 2.9$ TeV \cite{ATLAS}, \cite{CMS}.  So, most probably, the basic model will not be identified at this collider at all.  This problem must be attacked at the ILC.

Let us say a few words about the role of  beam polarizations $P_{e^+}, P_{e^-}$.   For the s-channel processes  the cross-section reads (see, for example,  eq.(3)in \cite{Godfrey}):
\be \label{SigmaPol} \sigma_{P_{e^+} P_{e^-}} = (1 - P_{e^+} P_{e^-})  [1 - P_{eff} A_{LR}]~ \sigma^{unpolarized}, \ee
where $A_{LR}$ is the left-right asymmetry and $P_{eff} = (P_{e^+} - P_{e^-})/(P_{e^+} P_{e^-} - 1)$ is the effective polarization. As we see, the  cross-section $\sigma_{P_{e^+} P_{e^-}} $is proportional to the unpolarized one.  The polarization dependent factors modify the effective luminosity for the process. This does not change qualitatively   the results discovered for unpolarized beams. 

 Now, let us compare the  obtained  results with the ones reported in the review \cite{Gulov2010}, where  the couplings $a^2$, $v^2_e$ and the mass $m_{Z'}$ were estimated within the data of the LEP1 and LEP2 experiments. The couplings  have been estimated at 1 - 2 $\sigma$ CL. This, in particular, means that the $Z'$ boson is Abelian one belonging  to the class covered by the relations \Ref{grgav}. The mass was estimated to be $\sim 1.1 - 1.4 $ TeV. This is in contrast to the results reported by the LEP Collaborations where no deviations from the SM  at the 2 $\sigma$ CL have been determined. In fact, the main goal of that investigations was searching for the $Z'$  particle at energies $\sim 100 - 200$ GeV. So, the observables introduced were constructed with accounting for the contact four fermion couplings, as it was done by the LEP Collaborations. The energy of the beams 500 GeV was also considered. However, the simple and specific  behavior of the factors $f_i(z)$ shown in Figs. 1, 2 was not determined. As a result, more complicated analysis was carried our and other observables for model-independent fitting of the $a^2$ and $v_f$ were used. Their efficiency is  sufficiently high. For instance, $\kappa_a^2 = 0.9587$ and  $\kappa_v =0.9533 $ (see Eqs.(33), (29) in \cite{Gulov2010}). The systematic error of the calculations was estimated to be  5 - 10 \%.   So, they also can be used in the analysis of experiments at the ILC and the results compared with obtained  on the base of the observables $A(E, m_{Z'})$ and $V(E, m_{Z'})$.

As
the present study shown,  information on the differential cross-sections with unpolarized beams is sufficient for determining important characteristics of the virtual $Z'$ state. Of course, it is of interest to consider  the case of polarized beams in more detail. It will be problem for the future.

 To complete, we would like to note that  the proposed observables are perspective for  consistent   analysis of  future  experiments at the ILC.

\section*{Acknowledgements}
The authors are grateful to  A.V. Gulov  and A.A. Pankov for fruitful discussions and suggestions and A. A. Kozhushko for the help in preparation of the package  for numeric calculations.

\section*{Appendix 1}
In this appendix, we adduce the expression  for the SM  differential cross-section and the factors $f_i(z, E)$ entering Eq.\Ref{5} and calculated in the improved Born approximation. To realize that we have
used the  packages  FeynArts \cite{FA}, FormCalc and LoopTools
\cite{FC} and Mathematica. The lepton masses are set to zero. For convenience, here we denote cosine of scattering angle as $x = z = \cos \theta$ and  introduce the standard notations: $s_W = \sin \theta_W, c_W = \cos \theta _W,$ where $ \theta_W $ is the Weinberg angle, $\alpha$ is a fine structure constant.

The differential cross-section reads
\begin{eqnarray} \label{A1}
\frac{\partial \sigma}{\partial x} = \sigma_{SM} + a^2 f_1^{\mu\mu}(x)  + v_e v_{\mu} f_2^{\mu\mu}(x) + a v_{e} f_3^{\mu\mu}(x) + a v_{\mu} f_4^{\mu\mu}(x).
\end{eqnarray}
In contrast to  Eq.\Ref{5} the factor $m_{Z'}^{-2}$ is incorporated in the functions. The cross-section is measured  in $GeV^{-2}$.

The SM part  is expressed in terms of the resonant functions $f_{Z}$ and $f_{ZE}$:
\begin{eqnarray}\label{A2}
\sigma_{SM} = \frac{\alpha^2 \pi}{32 s_W^4 c_W^4} \left\lbrace (1 + x^2)
\right.
\\
\times \left[ 4 s_W^4 c_W^4/E^2 + f_{ZE} (1 - 4 s_W^2 + 8 s_W^4)^2 + f_Z 2 s_W^2 c_W^2 (1 - 4 s_W^2)^2 \right]
\nonumber\\
+ \left. x \times \left[2 f_{ZE} (1 - 4 s_W^2)^2 + f_{Z} 4 c_W^2 s_W^2 \right]
\right\rbrace. \nonumber
\end{eqnarray}

The factors are expressed in terms of the resonant functions $f_{Z}$, $f_{Z'}$, $f_{ZE}$, $f_{ZZ'}$:
\begin{eqnarray}\label{A3}
f_1(x) = \frac{- \alpha}{64 s_W^4 c_W^4 m_{Z'}^4}\left\lbrace (1 + x^2)
\right.
\nonumber\\
\times \left[ f_{ZE} 4 c_W^2 s_W^2 m_Z^2 m_{Z'}^2 (1 - 4 s_W^2 + 8 s_W^4)
- f_{Z'} c_W^4 s_W^4 m_Z^4 (1 - 4 s_W^2)^2
\right.
\nonumber\\
\left.
- f_{ZE} f_{ZZ'} s_W^2 c_W^2 (m_{Z'}^2 + 2m_Z^2 (1 -4 s_W^2 + 8 s_W^4))^2 \right]
\nonumber\\
+ x \times \left[ f_Z 16 s_W^2 c_W^2 M_Z^2 M_{Z'}^2 - f_{Z'} 8 s_W^4 c_W^4 (m_Z^2 + m_{Z'}^2)^2 \right.
\nonumber\\
+ f_{ZE} 8 s_W^2 c_W^2 m_Z^2 m_{Z'}^2 (1 - 4 s_W^2)^2
\nonumber\\
\left. \left. - f_{ZE} f_{ZZ'} 2 s_W^2 c_W^2 (2 m_Z^2 + m_{Z'}^2) (1 - 4 s_W^2)^2 \right] \right\rbrace,
\end{eqnarray}

\begin{eqnarray}\label{A4}
f_3(x), f_4(x) = \frac{- \alpha}{64 s_W^4 c_W^4 m_{Z'}^4}\left\lbrace (1 + x^2)
\right.
\nonumber\\
\times \left[ (f_{Z} - f_{Z'}) 4 c_W^4 s_W^4 m_Z^2 m_{Z'}^2 (1 - 4 s_W^2)
\right.
\nonumber\\
\left.
+ f_{ZE} f_{ZZ'} s_W^2 c_W^2 (m_{Z'}^2 + 2 m_Z^2 (1 -4 s_W^2 + 8 s_W^4)) (1 - 4 s_W^2) m_{Z'}^2
\right.
\nonumber\\
\left.
+ f_{ZE} 2 s_W^2 c_W^2 m_Z^2 m_{Z'}^2 (-1 + 8 s_W^2 - 24 s_W^4 + 32 s_W^6)
\right]
\nonumber\\
x \times \left[
-f_{ZE} 4 s_W^2 c_W^2 m_Z^2 m_{Z'}^2 (1 - 4 s_W^2)
\right.
\nonumber\\
\left.\left.
+ f_{ZE} f_{ZZ'} 2 s_W^2 c_W^2 m_{Z'}^2 (2 m_Z^2 + m_{Z'}^2) (1 - 4 s_W^2)
\right] \right\rbrace,
\end{eqnarray}

\begin{eqnarray}\label{A6}
f_2(x) = \frac{- \alpha}{64 s_W^4 c_W^4 m_{Z'}^4}\left\lbrace (1 + x^2)
\right.
\nonumber\\
\times \left[ - f_{Z'} 4 s_W^4 c_W^4 m_{Z'}^4 - f_{ZE} f_{ZZ'} s_W^2 c_W^2 m_{Z'}^4 (1 - 4 s_W^2)^2 \right]
\nonumber\\
- \left. x \times 2 f_{ZE} f_{ZZ'} s_W^2 c_W^2 m_{Z'}^4 \right\rbrace.
\end{eqnarray}

The resonant functions are:
\begin{eqnarray}\label{A5}
f_{Z} = \frac{(4 E^2 - m_Z^2)}{(4 E^2 - m_Z^2)^2 + m_Z^2 \Gamma_Z^2}, \nonumber\\
f_{Z'} = \frac{(4 E^2 - m_{Z'}^2)}{(4 E^2 - m_{Z'}^2)^2 + m_{Z'}^2 \Gamma_{Z'}^2}, \nonumber\\
f_{ZE} = \frac{E^2}{(4 E^2 - m_Z^2)^2 + m_Z^2 \Gamma_Z^2},\nonumber\\
f_{ZZ'} = \frac{(4 E^2 - m_{Z'}^2) (4 E^2 - m_{Z}^2) + m_{Z'} \Gamma_{Z'} m_{Z} \Gamma_{Z}}{(4 E^2 - m_{Z'}^2)^2 + m_{Z'}^2 \Gamma_{Z'}^2}, \nonumber\\
\end{eqnarray}
where $\Gamma_{Z}, \Gamma_{Z'}$ are the widths of $Z$ and $Z'$ bosons.
\section*{Appendix 2}
In this appendix, we calculate a model-independent discovery reach for the mass $m_{Z'}$   based on the observable $V_{e \mu}$ Eq.\Ref{V} and data given in Table 3 for energy  500 GeV.  According to sect. 4, the observable reads
\be \label{DratV} R^{experim}_V(m_{Z'}) = \frac{V(E_1, m_{Z'})}{V(E_2, m_{Z'})} =
1 - \frac{\partial \ln V(E_1, m_{Z'})}{\partial E_1} \Delta E. \ee
As  $\Delta E$ we first take the
difference $\Delta E $ = 300 GeV between the  beam energies   $E_1 = $500 GeV and $E_2$ = 800 GeV planned for ILC experiments. The corresponding theoretical curve is the function $f_2^{\mu\mu}(E, z)$ Eq.\Ref{A6}.

Now,  we calculate the necessary constituents for the analysis (for more details see, for example, Ref. \cite{Pankov2013} where the process $e^+ e^- \to W^+ W^- $ is investigated). These are the integral $I^*_{SM}$ of the SM cross-section \Ref{A1} calculated  over the interval of interest   $- 1 \leq z \leq 0.5675 $. It gives a possibility for calculating in the SM the number $N^*_{SM} = I^*_{SM}~ L_{int}~ \epsilon_{\mu^+ \mu^-}$ of the processes $e^+ e^- \to \mu^+ \mu^-$  at a given integral luminosity $L_{int} = 500 fb^{-1}$. Here,  for simplicity, as the efficiency of the process reconstruction  we take $\epsilon_{\mu^+ \mu^-} = 0.95$ . We also neglect the systematic errors  which are to be much less than statistical ones. Since the beam energy is far from the resonance, we can put in Eq.\Ref{A5} $\Gamma_Z = \Gamma_{Z'} = 0 $. The observable  looks as follows
\be \label{Rf2} R_{f_2^{\mu\mu}}(m_{Z'}) = 1 - \frac{\int\limits_{-1}^{0.5675} (d f_2^{\mu\mu}(E, z)/d E)d z}{\int\limits_{-1}^{0.5675}  f_2^{\mu\mu}(E, z) dz}  |_{E = 500 GeV} \Delta E .\ee
To obtain the discovery reach we calculate the $\chi^2$ function
\be \label{Chi2} \chi^2 = \frac{(R_{f_2^{\mu\mu}}(m_{Z'}))^2}{(\delta R)^2_{SM}} \leq \chi^2_{min} + \chi^2_{CL}, \ee
where $(\delta R)_{SM} $ is the uncertainty  of the observable $R_{f_2^{\mu\mu}}(m_{Z'})$ calculated for  $N^*$, and find (at a chosen confidence level) the upper value of  $m_{Z'}$ below which the observable is reliable. The value of $\chi^2_{min}$ depends on the value of $\Delta E$. In the considered case we get $\chi^2_{min}$ = 0.2334.
In our analysis we choice $\chi^2_{CL} = 5.99$ that corresponds to 2 $\sigma$ CL. Then accounting for that for the function $R_v = (\Delta \sigma(E_1)/\Delta \sigma(E_2)$ the dispersion is calculated as
\be \label{dispersionR} ( \frac{\delta R_v}{R_v})^2 = ( \frac{\delta \Delta \sigma(E_1)}{\Delta \sigma(E_1)})^2 + ( \frac{\delta \Delta \sigma(E_2)}{\Delta \sigma(E_2)})^2\ee
and that $I^*_{SM} = 1.6586 \cdot 10^{-10}$ $ GeV^{-2}$, we estimate  the model-independent  discovery reach $m_{Z'}^{DRV} = 1.5 $ TeV. In the case of  $\Delta E$ = 10 GeV, $\chi^2_{min}$ = 269.84 and $m_{Z'}^{DRV}= 12 $ TeV.
\section*{Appendix 3}
One of  possibilities for determination of the $Z'$ discovery reach, $m_{Z'}^{DRA},$ is related with the observable $A(E, m_{Z'})$ \Ref{asyma2} and combining the results on estimating the coupling $a^2$ from  the data set of the LEP experiments. This value has been obtained as (see review \cite{Gulov2010} for details):
\be \label{aLEP} \frac{a^2}{m_{Z'}^2} = 1.97 \times 10^{-2} TeV^{-2}. \ee
By using this value and the results of section 3 and Appendix 1 we can construct the $\chi^2$ function:
\be \label{ChiA} \chi^2_A = \frac{(a^2 (\int\limits_{-1}^{-0.2} - \int\limits^{z*}_{-0.2})f_1^{\mu\mu}(z) d z )^2}{(\delta A)^2} \leq \chi^2_{min} +\chi^2_{CL}, \ee
where $(\delta A) $ is the uncertainty of the observable $A$, calculated for  $N^*$, and find (at a chosen confidence level) the upper value of  $m_{Z'}$ below which the observable is efficient. For this observable $\chi^2_{min}= 0. $ Expressing this function in terms of a number of particles we get
\be \label{ChiA1} \chi^2_A = \frac{(a^2 (\int\limits_{-1}^{-0.2} - \int\limits^{z*}_{-0.2})f_1^{\mu\mu}(z) d z )^2}{((\int\limits_{-1}^{-0.2} - \int\limits^{z*}_{-0.2})\sigma_{SM}(z) d z )^2} \frac{1}{N_1^{SM} + N_1^{SM}} \leq \chi^2_{CL}, \ee
where $N_1^{SM}, N_2^{SM}$ are the number of muon pairs in the backward and forward bins calculated in the standard model at a given luminosity 500 $fb^{-1}$ and reconstruction efficiency $\epsilon_{\mu\mu} = 0.95$, $\sigma_{SM}(z)$ is the differential cross section for the process calculated in the SM \Ref{A2}:
\bea \label{NN} && N_1^{SM} = | \int\limits_{-1}^{- 0.2}\sigma_{SM}(z) d z \times L_{int}~ \epsilon_{\mu^+ \mu^-}|, \nonumber \\
&& N_2^{SM} = | \int\limits_{-0.2}^{- 0.489}\sigma_{SM}(z) d z \times L_{int}~ \epsilon_{\mu^+ \mu^-}|. \eea
Assuming $\chi^2_{CL} = 5.99$ we obtain $m_{Z'}^{DRA} = 4.4 $ TeV.

\end{document}